\documentclass[aps,prb,onecolumn,footinbib,superscriptaddress]{revtex4-2}

\usepackage{amsmath}
\usepackage{amssymb}
\usepackage{siunitx}
\usepackage{graphicx}
\usepackage{xspace}
\usepackage{braket}
\usepackage{bm}
\usepackage{placeins}
\usepackage{leftidx}
\usepackage[usenames,dvipsnames]{color}
\usepackage[utf8]{inputenc}
\usepackage{natbib}
\usepackage[pdftex, colorlinks=true, urlcolor=blue, linkcolor=blue, citecolor=blue, pdfstartview=FitH, plainpages=false]{hyperref}
\usepackage{chemformula}
\usepackage{float}
\usepackage{todonotes}

\begin{document}

\title{Quantitative models for excess carrier diffusion and recombination in STEM-EBIC experiments on semiconductor nanostructures}
\date{\today}

\author{T. Meyer}
\affiliation{4th Institute of Physics –Solids and Nanostructures, Georg-August-University Goettingen, Friedrich-Hund-Platz 1, Göttingen 37077, Germany\\}
\affiliation{Institute of Materials Physics, Georg-August-University Goettingen, Friedrich-Hund-Platz 1, Göttingen 37077, Germany}
\email{tmeyer@uni-goettingen.de}
\author{C. Flathmann}
\affiliation{4th Institute of Physics –Solids and Nanostructures, Georg-August-University Goettingen, Friedrich-Hund-Platz 1, Göttingen 37077, Germany\\}
\affiliation{Research Center Future Energy Materials and Systems, Ruhr University Bochum, Universitätsstr. 150, Bochum 44801, Germany\\
Faculty of Physics and Astronomy, Ruhr University Bochum, Universitätsstr. 150, Bochum 44801, Germany\\}
\author{D.A. Ehrlich}
\affiliation{4th Institute of Physics –Solids and Nanostructures, Georg-August-University Goettingen, Friedrich-Hund-Platz 1, Göttingen 37077, Germany\\}
\author{P. Paap-Peretzki}
\affiliation{4th Institute of Physics –Solids and Nanostructures, Georg-August-University Goettingen, Friedrich-Hund-Platz 1, Göttingen 37077, Germany\\}
\author{J. Lindner}
\affiliation{Institute of Materials Physics, Georg-August-University Goettingen, Friedrich-Hund-Platz 1, Göttingen 37077, Germany}
\author{C. Jooß}
\affiliation{Institute of Materials Physics, Georg-August-University Goettingen, Friedrich-Hund-Platz 1, Göttingen 37077, Germany}
\author{M. Seibt}
\affiliation{4th Institute of Physics –Solids and Nanostructures, Georg-August-University Goettingen, Friedrich-Hund-Platz 1, Göttingen 37077, Germany\\}


\begin{abstract}

The increased complexity and reduced size of (opto-)electronic devices demands for quantitative descriptions of excess carrier transport and recombination via various mechanisms. In addition, experimental methods capable of resolving carrier dynamics on the nanometer scale are required. In this paper, we present a quantitative model of a confined geometry including recombination at two surfaces, which is very generic for electron beam induced current measurements in a scanning transmission electron microscope -- a method which offers atomic scale spatial resolution.
The model is based on analytical considerations as well as finite element simulations and underlying assumptions are subjected to an in-depth discussion.
Finally, the successfull application to experimental data
obtained on the complex oxide SrTi$_{0.995}$Nb$_{0.005}$O$_3$ demonstrates the practicality and robustness of the approach, which enables the precise determination of its bulk diffusion length of $L=10.2\pm0.1\,$nm.

\end{abstract}

\maketitle


\section{Introduction}

Designing and understanding the functionality of nanometer scale devices is key to several (opto-)electronic applications. Examples include solar to electric energy conversion in ultrathin photovoltaic devices \cite{massiot2020progress}, light emission from nitride quantum wells \cite{sheen2022highly}, or -- among other fascinating fundamental effects -- photon detection in two-dimensional transition metal dichalcogenides \cite{wang2012electronics}.
Recently, metal halide perovskites gained particular interest due to their potential application in thin film solar cells \cite{green2014emergence} and light-emitting diodes \cite{fakharuddin2022perovskite}, although, improving their stability \cite{du2024improving} and avoiding usage of lead \cite{miyasaka2020perovskite,hoffmann2020fabrication} remain major challenges.

Fundamentally, one can separate the optimization of related functional devices into (i) challenges that arise due to bulk properties and (ii) those related to device architectures.
A prototypical example for (i) in the context solar cells is given by the prominent Shockley-Queisser limit \cite{shockley32detailed} yielding a material-specific maximal value for the power conversion efficiency.
Aligned with the approach of studying single material properties, the effect of bulk defects on photovoltaic performance was studied in classical semiconductors \cite{queisser1998defects} as well as halide perovskites \cite{ball2016defects}. Furthermore, systems like transition metal oxides in which hot charge carriers can be harvested were studied and hypothesized to exhibit carrier stabilization due to charge \cite{kressdorf2020room} and orbital ordering \cite{kressdorf2021orbital}.
Concerning aspect (ii), extensive efforts were made to improve the behavior of surfaces and interfaces inevitably included in functional devices:
While the surface recombination velocity of solar cells was drastically reduced by passivation \cite{schmidt2018surface}, the recombination at contact interfaces was optimized by reducing the contact area \cite{blakers2019development} and adding passivation as well as carrier selective layers \cite{titova2018implementation,flathmann2023composition}.

Besides testing the overall performance of macroscopic devices, several microscopic (photo-)electrical characterization methods are available including surface probe techniques like Kelvin probe force microscopy \cite{vishwakarma2018direct}, scanning capacitance microscopy \cite{matsumura2014characterization}, or scanning spreading resistance microscopy \cite{eyben2011development}. Furthermore, photon probes are employed during photoluminescence \cite{trupke2012photoluminescence} and light/laser beam induced current \cite{moralejo2010lbic} measurements as well as electron probes in their respective counterparts called cathodoluminescence \cite{coenen2017cathodoluminescence} and electron beam induced current (EBIC) \cite{leamy1982charge}.
The latter -- classically conducted in a scanning electron microscope (SEM-EBIC) -- combines high spatial resolution and relatively straight-forward interpretation of the signal as excess charge carrier collection efficiency with the possibility of tuning the electron beam energy and thus the contribution of bulk and surface effects.

Based on the analytical solutions found in \cite{van1955injected} for the excess charge carrier concentration in a neutral semiconductor terminated by an uncharged surface, several strategies were developed to disentangle the effect of bulk and surface recombination: 
Accounting for the interaction volume of the electron beam with the sample and tailored geometries used in SEM-EBIC experiments, relations between experimentally observed profile decay lengths and the bulk diffusion length were reported \cite{hackett1972electron,donolato1983evaluation,donolato1999reconstruction,luke1985quantification}. Furthermore, expressions for point generations at a certain depth \cite{jastrzebski1975application} or close to the surface \cite{berz1976theory,ong1994direct} were suggested and numerical evaluation of infinite series \cite{tan2013study} and random walk on spheres based stochastic simulations \cite{shalimova2018random} were used to calculate excess charge carrier collection efficiency profiles.

To further improve the spatial resolution -- limited by the electron interaction volume in SEM-EBIC -- electron transparent samples were prepared and successfully investigated using scanning transmission electron microscopy (STEM) in \cite{cabanel2006low}.
Owing to the small lateral extent of the electron beam, atomically resolved signals can be obtained \cite{meyer2020structural} and several nanostructures were successfully investigated using STEM-EBIC including complex oxide and gallium arsenide heterojunctions \cite{meyer2019high,zutter2021mapping}, silicon homojunctions \cite{conlan2021electron,schneider2025stem}, as well as halide perovskite thin films \cite{duchamp2020stem}.
Importantly, the same experimental setup can be employed to measure currents induced by the emission of secondary electrons (SE), which is also referred to as SEEBIC, yields atomic scale signals as well \cite{mecklenburg2019electron,dyck2023direct}, and can be used for local thermometry \cite{hubbard2023emission} or visualization of nanoparticle morphologies \cite{vlasov2023secondary}. However, the SE signal is much smaller than currents produced by excess charge carriers and thus contributes only to a small background in STEM-EBIC profiles across rectifying junctions.

In this paper, we combine analytical considerations with numerical simulations using the finite element method (FEM) to establish a quantitative formalism describing the interplay of diffusion as well as bulk and surface recombination of excess charge carriers in a finite, electron transparent semiconductor after a stripe-shaped generation. This enables us to measure nanometer bulk diffusion lengths in STEM-EBIC experiments. The established formalism is applied to experimental data obtained on the complex oxide SrTi$_{0.995}$Nb$_{0.005}$O$_3$ yielding a value of $10.2\pm0.1\,$nm and the model assumptions as well as possible extensions are carefully discussed.

\section{Results and Discussion}

\begin{figure}[h]
  \includegraphics[width=\linewidth]{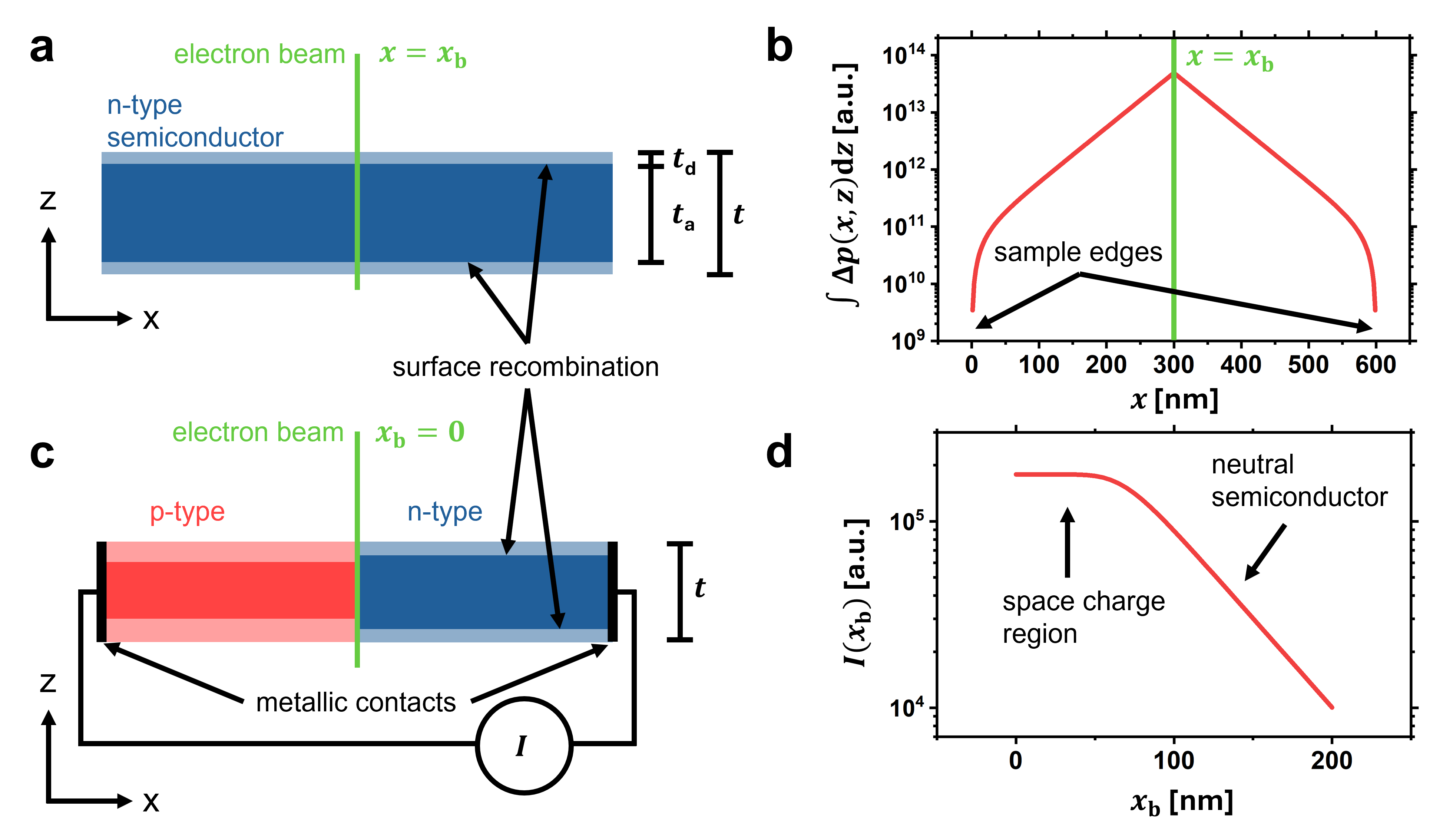}
  \caption{Overview of modeled geometries: (a) A neutral n-type semiconductor of finite thickness $t$ is excited by a high-energy electron beam causing a stripe shaped generation of excess charge carriers. The material consists of an active interior area with thickness $t_\text{a}$ and dead surface layers with thickness $t_\text{d}$ and surface recombination occurs at the edges of the active layer. (b) Resulting profile of the excess hole concentration $\Delta p$ integrated along the $z$ axis. Deviations from an exponential decay at the sample edges are caused by the boundary condition $\Delta p=0$. (c) Extended geometry including a junction to a p-type semiconductor (with possibly different dead layer thickness) and metallic contacts to probe the short-circuit current caused by the electron beam, i.e., the STEM-EBIC signal. (d) Resulting STEM-EBIC profile with beam positions located in the neutral and space charge region of the n-type material.}
  \label{fig:geometry}
\end{figure}

Two different geometries are modeled in this study to investigate the effect of surface recombination in electron transparent samples: 
Firstly, a homogeneous n-type semiconductor of thickness $t$, in which a stripe generation of excess charge carriers along the $z$-direction is assumed as shown in Figure \ref{fig:geometry}a. Secondly, a pn junction as illustrated in Figure \ref{fig:geometry}c resulting in a short circuit current through metallic contacts at the semiconductor edges, i.e., a STEM-EBIC signal.
In both cases, it is assumed that the materials consist of active interior regions with thickness $t_\text{a}$ as well as electronically inactive surface layers with thickness $t_\text{d}$. The latter are also referred to as dead layers and were reported before in the context of electron holography \cite{rau1999two,twitchett2002quantitative} as well as STEM-EBIC \cite{peretzki2017low} to explain vanishing experimental signals at finite sample thicknesses. A more detailed discussion about their possible microscopic nature is given later and surface recombination is assumed to occur at the interface between active and dead layers.
The homogeneous geometry in Figure \ref{fig:geometry}a is used to study the total number of excess holes caused by the electron beam as well as decay lengths in excess hole concentration profiles integrated along the $z$ axis as shown in Figure \ref{fig:geometry}b. Obtained formalisms can analogously be applied to excess electrons in p-type semiconductors as well. In addition, the pn junction presented in Figure \ref{fig:geometry}c is evaluated for varying electron beam positions to determine the decay length in the neutral semiconductor region of STEM-EBIC profiles as shown in Figure \ref{fig:geometry}d.
The results are structured in three subsections containing analytical considerations, numerical findings, and the application of resulting models to experimental data.

\subsection{Analytical description of excess charge carriers and effective diffusion lengths}

The interplay of electrons, holes, and the electrostatic potential in a doped semiconductor is described by the van-Roosbroeck set of differential equations \cite{van1950theory}:
\begin{align}
    -\vec{\nabla}\left(\epsilon_0\epsilon_\text{r}\vec\nabla\phi\right)=e\left(N_\text{d}-N_\text{a}+p-n\right)\label{eq:poisson}\,,\\
    0=\dot{n}=-\vec\nabla\left(\mu_\text{e}n\vec\nabla\phi-D_\text{e}\vec\nabla n\right)+g-r\label{eq:n}\,,\\
    0=\dot{p}=-\vec\nabla\left(-\mu_\text{h}p\vec\nabla\phi-D_\text{h}\vec\nabla p\right)+g-r\label{eq:p}\,.
\end{align}
Here, $n$ and $p$ are the electron and hole concentrations, $\phi$ is the electrostatic potential, $\epsilon_0$ and $\epsilon_\text{r}$ are the vacuum and relative permittivity, $N_\text{d}$ and $N_\text{a}$ are the concentrations of ionized donator and acceptor dopants, $\mu_\text{e}$ and $\mu_\text{h}$ are the electron and hole mobilities, $D_\text{e}$ and $D_\text{h}$ are the electron and hole diffusivities, $g$ and $r$ are the generation and recombination functions (being equal for electrons and holes). Setting the time derivatives $\dot{n}$ and $\dot{p}$ to zero corresponds to the stationary case.

The effect of enhanced excess charge carrier recombination due to a surface defect level can be modeled including a Shockley-Read-Hall recombination term for all surface positions \cite{haney2016depletion}:
\begin{align}
 \hat{n}\cdot D\vec\nabla p=s\frac{np-n_i^2}{n+n_1+p+p_1}\,.\label{eq:SRH_recombination}
\end{align}
Here, $\hat{n}$ is the surface normal vector, $s$ the surface recombination velocity, $n_\text{i}$ the intrinsic charge carrier concentration and the electron and hole concentrations $n_1$ and $p_1$ correspond to the values expected if the electrochemical potential is located at the defect level.
Furthermore, $D=(n_\text{eq}+p_\text{eq})/(n_\text{eq}/D_\text{h}+p_\text{eq}/D_\text{e})$ is the ambipolar diffusivity related to the equilibrium concentrations of electrons $n_\text{eq}$ and holes $p_\text{eq}$.
In case of an n-type material ($n\gg p$), weak injection ($n\approx n_\text{eq}$), and a deep level defect ($n\gg n_1+p_1$), Equation (\ref{eq:SRH_recombination}) can be approximated as 
\begin{align}
    \hat{n}\cdot D\vec\nabla p\approx s\Delta p\,,\label{eq:surface_recombination}
\end{align}
with the excess charge carrier concentration $\Delta p=p-p_\text{eq}$.

A general solution of Equations (\ref{eq:poisson})-(\ref{eq:p}) for arbitrary geometries remains elusive, however, simplified cases can be evaluated analytically:
In \cite{van1955injected}, a field-free, single material n-type semiconductor was considered in a semi-infinite geometry including point, stripe, and planar generation functions and explicit expressions for the excess hole concentrations was found using Green's functions.
Based on these results, the total number of excess holes in the semi-infinite case and for a point generation with generation function $g=G\delta\left(x,y,z-\xi\right)$ was found in \cite{jastrzebski1975application} to be:
 \begin{align}
     \Delta p^\text{tot}_\text{semi-inf}(\xi) = \frac{GL^2}{D}\left(1-\frac{s}{s+1}\exp(-\xi/L)\right)=\Delta p^\text{tot}_{\infty}\left(1-\frac{s}{s+1}\exp(-\xi/L)\right)\,.\label{eq:delta_p_semi_inf}
 \end{align}
Here, $G$ denotes the generation strength, $L$ the bulk diffusion length for holes, and $\xi$ the generation distance from the surface. Furthermore, $\Delta p^\text{tot}_{\infty}=GL^2/D$ corresponds to the total number of excess holes for a recombination-free surface (or an infinite semiconductor). We emphasize that $\Delta p^\text{tot}_\text{semi-inf}$ and $\Delta p^\text{tot}_{\infty}$ -- obtained by integrating the excess hole concentration over all spatial coordinates -- are dimensionless.

By accounting for a second surface with an additional reduction term in Equation (\ref{eq:delta_p_semi_inf}) and an adopted boundary condition analogous to Equation (\ref{eq:surface_recombination}) as shown in \cite{peretzki2019implementation}, an analytical expression for the total number of excess holes can be derived for a finite geometry with (active) thickness $t_\text{a}=t-2t_\text{d}$. Averaging the result for all generation depths contained in the active sample region, i.e., assuming a stripe generation of excess charge carriers yields the following expression:
\begin{align}
    \Delta p^\text{tot}=\Delta p^\text{tot}_{\infty}\left(1-\frac{\frac{2L}{t_\text{a}}}{\frac{D}{sL}+\coth{\left(\frac{t_\text{a}}{2L}\right)}}\right)\,,\quad t_\text{a}=t-2t_\text{d}.\label{eq:delta_p_finite}
\end{align}
Following the approach in \cite{jastrzebski1975application} to define a global, effective lifetime satisfying $\tau^\text{eff}/\tau=\Delta p^\text{tot}/\Delta p^\text{tot}_{\infty}$ (with $\tau$ being the bulk lifetime) leads to an effective diffusion length of
\begin{align}
    L^\text{eff}_\tau=\sqrt{D\tau^\text{eff}}=L\cdot\sqrt{1-\frac{\frac{2L}{t_\text{a}}}{\frac{D}{sL}+\coth{\left(\frac{t_\text{a}}{2L}\right)}}}\,,\quad t_\text{a}=t-2t_\text{d}.\label{eq:L_tau}
\end{align}
The subscript $\tau$ indicates that the effective diffusion length was calculated via the effective lifetime.

We emphasize that the validity of both the ansatz for an additional surface term leading to Equation (\ref{eq:delta_p_finite}) as well as the assumption that a global effective lifetime can be used to derive Equation (\ref{eq:L_tau}) remains unclear at this point and will be investigated in the following sections, while Equation (\ref{eq:delta_p_semi_inf}) is a direct consequence of the solutions presented in \cite{van1955injected}.

\subsection{Numerical simulation results and an empirical correction}

To check the validity of Equation (\ref{eq:delta_p_finite}) and (\ref{eq:L_tau}), the two geometries in Figure \ref{fig:geometry}a and \ref{fig:geometry}c are modeled numerically and Equations (\ref{eq:poisson})-(\ref{eq:p}) are solved via FEM using the surface recombination term given in Equation (\ref{eq:SRH_recombination}). A midgap defect level is assumed, i.e., the defect level is set to the center of the band gap, and the inactive, dead layers are omitted in the numerical model.
More details about the simulation parameters and boundary conditions are given in the supporting information.
Exemplary profiles of resulting excess hole concentrations and STEM-EBIC signals were already presented in Figure \ref{fig:geometry}b and \ref{fig:geometry}d and exhibit following behavior: 
The excess hole concentration profile decays exponentially with increasing distance from the electron beam position (except in the vicinity of the sample edges, where $\Delta p=0$ is enforced by the boundary conditions). The STEM-EBIC profile, however, decays only exponentially in the neutral semiconductor region (and well-separated from the contacts), while a saturation is observed in the space charge region involving finite electric fields. We emphasize that depending on the recombination mechanisms and electric field strength, sharp maxima can be observed in the space charge region as well, which is the case for the experimental data shown in the following subsection.
To reliably extract the related decay lengths, a tailored criterion to objectively define a suitable fitting regime excluding space charge region and edge effects is used, which is described in detail in the supporting information. The decay length obtained from the excess hole concentration is denoted by $L_{\Delta p}$ and the decay length of STEM-EBIC profiles by $L_I$.

\begin{figure}
  \includegraphics[width=\linewidth]{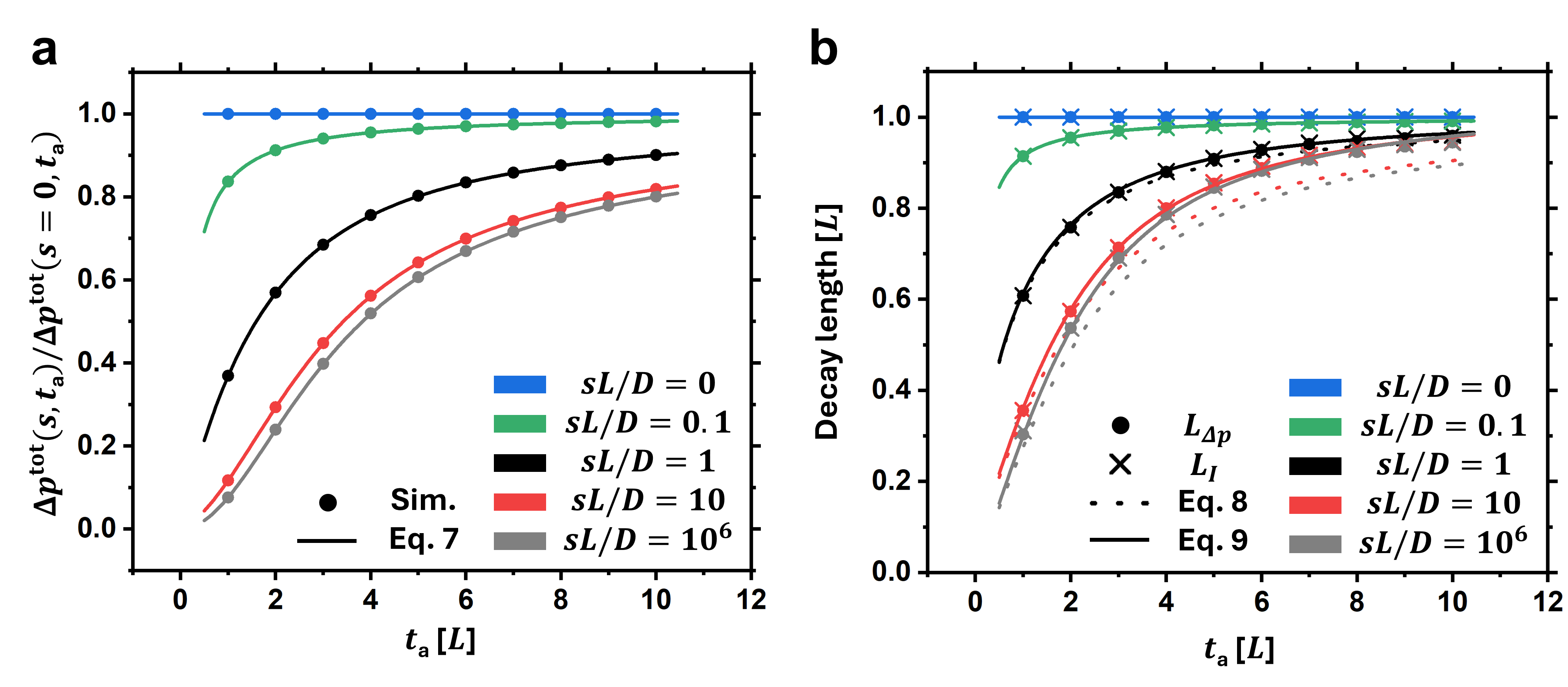}
  \caption{Comparison between numerical simulation results and model equations for different values of the surface recombination velocity $s$ as a function of active thickness $t_\text{a}$: (a) shows the total number of excess holes $\Delta p^\text{tot}$ obtained via FEM for the geometry in Figure \ref{fig:geometry}a (points) and the analytical prediction by Equation (\ref{eq:delta_p_finite}) (lines). (b) shows the decay lengths $L_{\Delta p}$ and $L_{I}$ obtained from exponential fits of the excess hole concentration integrated along $z$ (points, model shown in Figure \ref{fig:geometry}a) as well as the STEM-EBIC profiles (crosses, model shown in Figure \ref{fig:geometry}c). In addition, the predictions by Equations (\ref{eq:L_tau}) and (\ref{eq:L_empirical}) are shown as dashed and solid lines.}
  \label{fig:simulation}
\end{figure}

The resulting values for $\Delta p^\text{tot}$, $L_{\Delta p}$, and $L_{I}$ are plotted as points in Figure \ref{fig:simulation} for different thicknesses and surface recombination velocities enabling a comparison to the analytical equations included as lines:
In fact, the FEM results of $\Delta p^\text{tot}$ in Figure \ref{fig:simulation}a perfectly agree with Equation (\ref{eq:delta_p_finite}) proving that the additional surface term used to derive it from the semi-infinite case correctly models the finite geometry.
Furthermore, $\Delta p^\text{tot}$ decreases as expected monotonously with increasing $s$ or decreasing $t_\text{a}$. 
Qualitatively, the same trend is visible for the two decay lengths $L_{\Delta p}$ (points) and $L_{I}$ (crosses) in Figure \ref{fig:simulation}b, which are virtually identical. Importantly, the observation $L_{\Delta p}=L_{I}$ confirms that STEM-EBIC profiles in a neutral, extrinsic semiconductor directly probe the excess carrier diffusion and justifies the definition of a single effective diffusion length by the common value.
A quantitative match between the simulation results with Equation (\ref{eq:L_tau}) (dashed lines), however, is only observed for small values of $s$ or $t_\text{a}$.
Thus, considering the total number of excess charge carriers and effective diffusion lengths leads to different values of corresponding effective lifetimes, which were assumed to be identical when deriving Equation (\ref{eq:L_tau}) from (\ref{eq:delta_p_finite}).
Lacking an analytical solution, we propose an empirical correction accounting for the systematic underestimation of the effective diffusion length by Equation (\ref{eq:L_tau}), which is given by the following expression:
\begin{align}
    L^\text{eff}_\text{emp}=L^\text{eff}_\tau
    +0.0666\cdot\frac{(\frac{t_\text{a}}{L})^2}{1+\frac{t_\text{a}}{L}}
    \cdot\frac{\frac{sL}{D}}{1+\frac{sL}{D}}\cdot\left(L-L^\text{eff}_\tau\right)\,,\quad t_\text{a}=t-2t_\text{d}.\label{eq:L_empirical}
\end{align}
The last factor of the correction term $L-L^\text{eff}_\tau$ is motivated by the fact that all data points in Figure \ref{fig:simulation}b are located between $L$ and $L^\text{eff}_\tau$, while the two central terms are chosen to guarantee that the correction term vanishes for small values of $s$ or $t_\text{a}$. The quadratic dependence of $t_\text{a}/L$ in the nominator of the second term is found to describe the data points slightly better than a linear term and the prefactor of $0.0666$ is found by minimizing the mean square difference between simulated results and Equation (\ref{eq:L_empirical}). 
Importantly, spot-checked changes of the used numerical parameters such as mesh and model sizes, the defect energy level, the bulk recombination parameter, doping levels, or mobilities lead to results identical to those in Figure \ref{fig:simulation} implying that the empirically found prefactor does not depend on material properties, but reflects the underlying geometry.
Even though the empirical solution is not derived mathematically, its high accuracy in describing the results obtained in the FEM simulations across the entire range of relevant values for $s$ and $t_\text{a}$ suggests, that it can be readily used to extract quantitative values for $L$ and $s$ in STEM-EBIC experiments.

\subsection{Experimental quantification of bulk diffusion lengths and surface recombination velocities}

\begin{figure}
  \includegraphics[width=\linewidth]{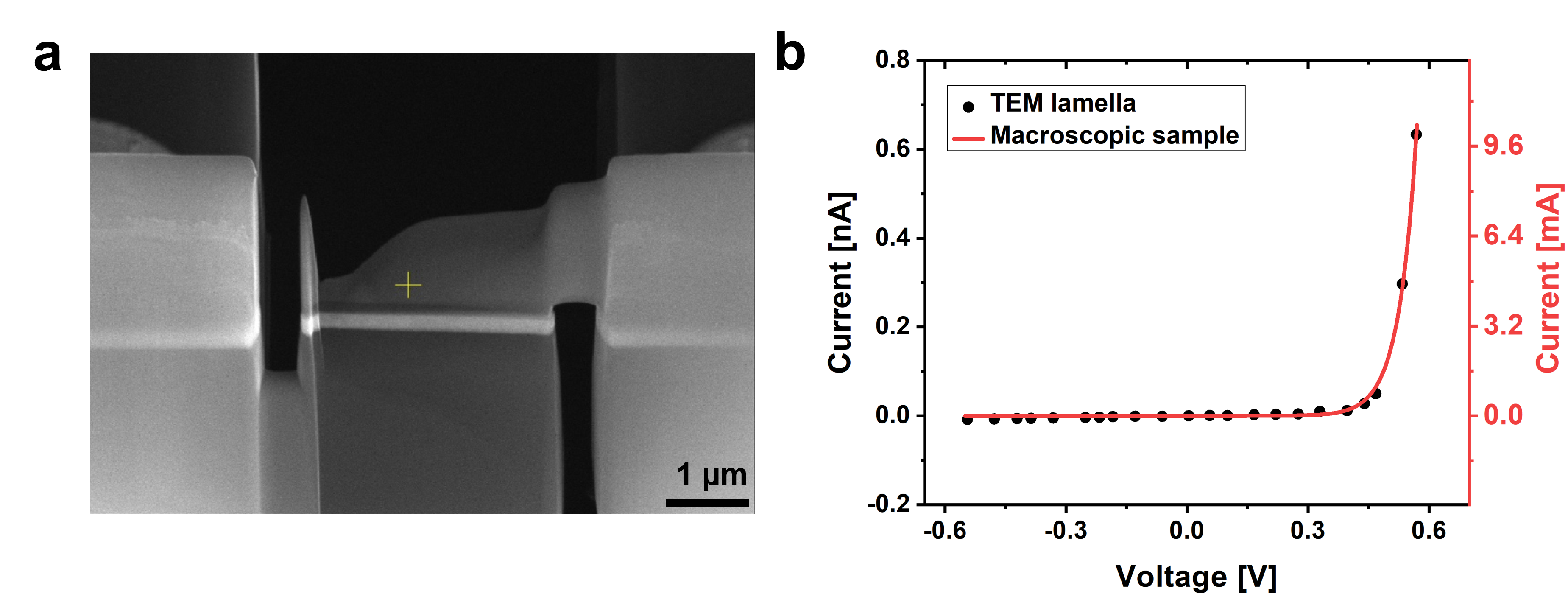}
  \caption{(a) SEM overview of the lamella extracted from an RP-PCMO-STNO junction and mounted on a MEMS chip. Vertical cuts were applied to prevent short circuits across the junction. (b) Comparison of the current-voltage characteristics of the lamella (black points) and the macroscopic sample (red line).}
  \label{fig:sample_prep}
\end{figure}

\begin{figure}
  \includegraphics[width=\linewidth]{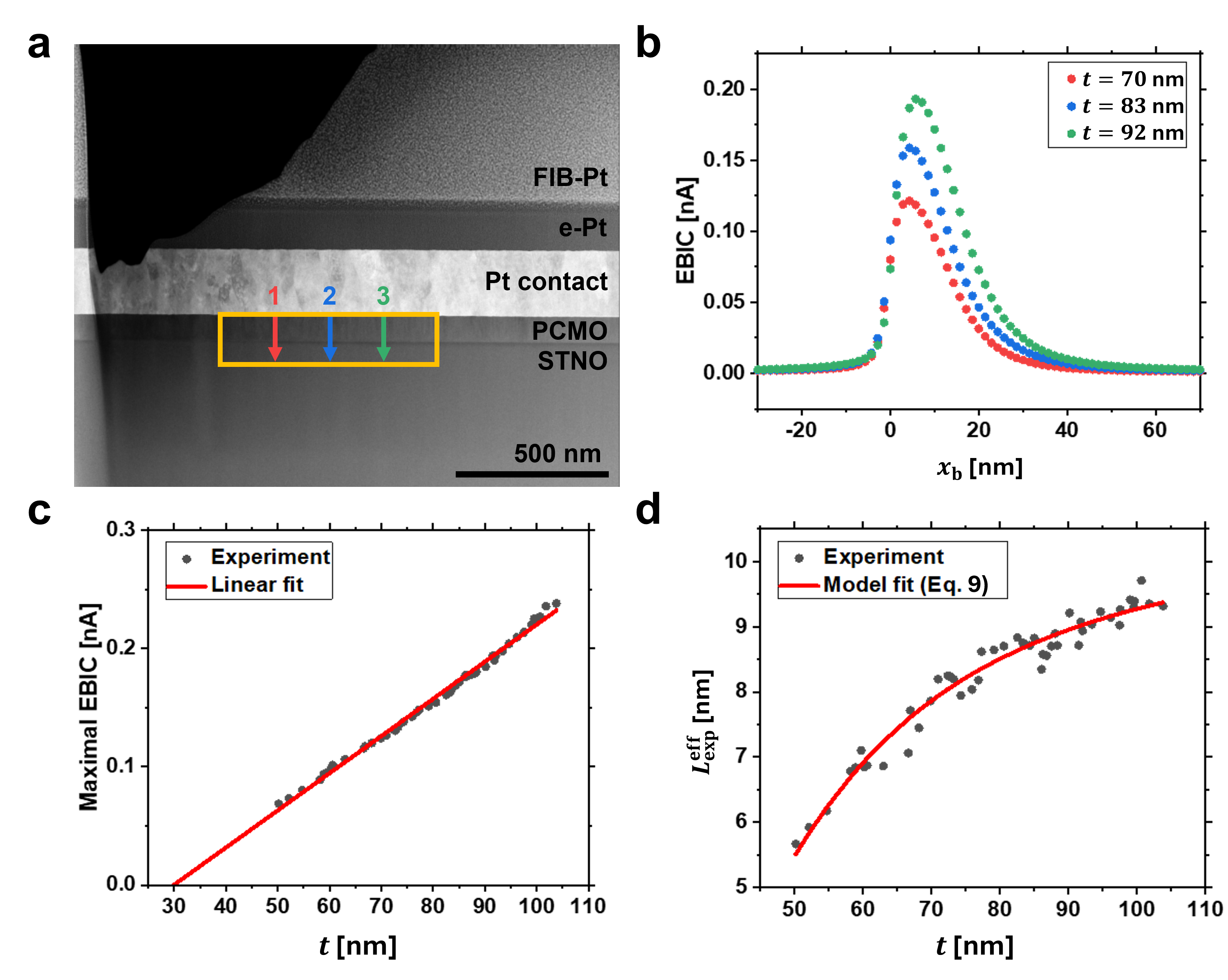}
  \caption{(a) Annular dark-field STEM image showing all layers of the junction and the front contact. The region marked by the yellow rectangle was analyzed with vertical STEM-EBIC scans and contains an intentional thickness gradient in the horizontal direction. The three arrows indicate the profiles along which the current is shown in (b). (c) and (d) show the thickness dependence of the line profiles' maximal STEM-EBIC signal and decay length in the n-type STNO substrate (points). While (c) includes a linear regression of the data points, a model fit to Equation (\ref{eq:L_empirical}) is given in (d) (lines).}
  \label{fig:experiment}
\end{figure}

In this section, the formalisms presented in the previous two subsections are applied to experimental data obtained on a heterojunction formed by a Ruddlesden-Popper Pr$_{0.5}$Ca$_{1.5}$MnO$_4$ (RP-PCMO) thin film on a SrTi$_{0.995}$Nb$_{0.005}$O$_3$ (STNO) substrate. The thin film is grown by ion beam sputtering and an electron transparent sample is extracted using focused ion beam (FIB). More details about the macroscopic and TEM sample preparation are given in the experimental section and an SEM overview of the final lamella is shown in Figure \ref{fig:sample_prep}a. 
Importantly, the latter is thinned in a wedge shape such that an intentional thickness gradient is present from left to right.
The current-voltage characteristic of the TEM lamella shows strong rectifying behavior and is perfectly consistent with the data obtained on the macroscopic sample as demonstrated in Figure \ref{fig:sample_prep}b.
To investigate the thickness dependence of STEM-EBIC signals across the heterojunction, vertical profiles are collected in the region marked by the yellow rectangle in the annular dark-field STEM image in Figure \ref{fig:experiment}a.
Due to the wedge shape of the lamella, different horizontal positions correspond to different local thicknesses and details how the latter is determined using electron energy loss spectroscopy (EELS) are given in the supporting information.
Three exemplary profiles marked by arrows in Figure \ref{fig:experiment}a are plotted in Figure \ref{fig:experiment}b showing clear STEM-EBIC signals in the vicinity of the junction. 
As reported previously for related manganite thin films \cite{peretzki2017low,meyer2019high}, the signal in the RP-PCMO is low due to the short lifetime of excess carriers and thus the discussion will focus on the n-type STNO substrate.
Two main observation can be inferred from the profiles: Firstly, the maximum current increases with the sample thickness, which is expected due to 
the longer interaction path of the electrons leading to more excess carrier generation. Secondly, the increased signal width in the STNO indicates higher effective diffusion lengths for higher thicknesses.
For further illustration, the maximal EBIC values as well as the effective diffusion lengths are shown as a function of thickness in Figure \ref{fig:experiment}c+d. The latter are extracted analogously to those from simulated profiles as documented in the supporting information.
The maximal EBIC in Figure \ref{fig:experiment}c shows a linear dependence in the investigated thickness regime and the extrapolated linear fit intersects the thickness axis at $30\pm1$\,nm. Since the maximum position is obtained in the substrate, this indicates a dead layer thickness of $15.0\pm0.5$\,nm for STNO. 
Furthermore, the effective diffusion length given in Figure \ref{fig:experiment}d increases -- as expected -- with the thickness and the least-squares model fit to Equation
(\ref{eq:L_empirical}) describes the experimental data remarkably well. The obtained fit parameters are $L=10.2\pm0.2\,$nm, $s/D=1.6\times10^5\pm8\times10^{10}\,$nm$^{-1}$, and $t_\text{d}=15\pm5\,$nm.
The errors are calculated from the covariance matrix of the least-squares fit and the fact that the value for $s/D$ is five orders of magnitude below its estimated error reflects that the model equation is virtually independent of $s/D$ for the obtained optimal parameters.
This finding perfectly aligns with the numerical results in Figure \ref{fig:simulation} showing that the effective diffusion length hardly changes for values of $sL/D>10$. Therefore, a second fit assuming the limiting case $s=\infty$ is performed yielding the values $L=10.2\pm0.1\,$nm and  $t_\text{d}=15.0\pm0.3\,$nm.
Consistently, the values are identical to those obtained including three fit parameters and solely the estimated errors are decreased.
Remarkably, the value for the dead layer thickness equals that one extracted from the intersect of the EBIC maximum in Figure \ref{fig:experiment}c underlining the model's overall accuracy.

\section{Conclusion and Outlook}

Quantitative modeling of excess charge carrier dynamics in confined semiconductor geometries is of high relevance to describe non-equilibrium states in (opto-)electronic devices.
In this paper, analytical expressions describing a thin, electron transparent sample and including surface recombination were revised and tested by finite element simulations as well as experimental data resulting in a quantitative match and successful determination of the bulk diffusion length $L=10.2\pm0.1\,$nm for SrTi$_{0.995}$Nb$_{0.005}$O$_3$. Furthermore, it was shown that the surface recombination velocity is virtually infinity after FIB preparation and dead layers with a thickness of $t_\text{d}=15.0\pm0.3\,$nm reduce the effective sample thickness in STEM-EBIC experiments.

The presented findings have several important implications related to semiconductor modeling in general as well as EBIC experiments in particular: 
Firstly and most importantly, the results clearly define suitable strategies to extract bulk diffusion lengths on the order of several nanometers to a few hundred nanometers with STEM-EBIC: For values up to a few tens of nanometers, a constant sample thickness around ten times the bulk diffusion length or higher can be used leading to deviations of the effective value by less than 10\,\%. In contrast, for larger values, the thickness-dependent approach applied in this paper can be used to extrapolate the value of the bulk diffusion length or -- if the spatial resolution is sufficient -- SEM-EBIC can be employed.
Secondly, we have shown that an analytical expression for the total number of excess charge carriers $\Delta p^\text{tot}$ can be obtained by incorporating several reduction terms each corresponding to individual surfaces.
This concept might be transferrable to more complex geometries like faceted nanowires or quantum dots. 
Lastly, the perfect match between the resulting three-dimensional analytical description of $\Delta p^\text{tot}$ and the two-dimensional simulation results implies that a reduction of the dimensionality of translational invariant geometries is generally justified. The approach to reduce the dimensionality was taken further by introducing an effective diffusion length and thus an effectively one-dimensional description of STEM-EBIC profiles in the neutral semiconductor region. Importantly, the latter could not be achieved by using the effective carrier lifetime deduced from the total number of excess carriers as previously assumed in \cite{jastrzebski1975application}, but required an empirical correction backed by the FEM simulations.

Future work should focus on formalisms including surface charges or the electric field inside the space charge region. The former are expected to be strongly linked to the dead layers observed in experiments and the latter could be addressed experimentally by mapping the electric field with electron holography \cite{anada2020direct,denaix2023inversion,ccelik2024simple,lindner2024reconstruction} or four-dimensional STEM \cite{wu2023toward,toyama2023real,wartelle2025sub,flathmann2025sequential} -- methods that can also be used to correlate the observed electronic properties with phase transitions related to long-living hot carriers \cite{meyer2021phase,flathmann2024relationship}.
Furthermore, an analytical explanation of the used empirical correction 
is desirable.

In short, we have demonstrated for the first time how nanometer scale diffusion lengths can be precisely determined using STEM-EBIC and the concepts used in this paper will largely help to establish quantitative models describing excess charge carrier dynamics in nanoscale (opto-)electronic devices.


\section{Experimental Section}

\subsection{Finite element simulations}

Two-dimensional finite element simulations were conducted with COMSOL Multiphysics v6.1.0.357 using a stationary solver and including the electric potential as well as the electron and hole concentrations as variables. The minimal and maximal mesh size was set to 0.02\,nm and 2\,nm and all relevant parameters, used boundary conditions, and the definition of the STEM-EBIC probe are given in the supporting information.

\subsection{Macroscopic sample preparation}

An 80\,nm thick Pr$_{0.5}$Ca$_{1.5}$MnO$_4$ film was epitaxially grown on a single-crystalline, doubleside-polished, (110)-oriented SrTi$_{0.995}$Nb$_{0.005}$O$_3$ substrate via ion beam sputtering using pressures of $p_\text{Ar}=3\times10^{-4}$\,mbar (beam neutralizer),
$p_\text{Xe}=1\times10^{-4}$\,mbar (sputter gas), and 
$p_\text{O}=1.4\times10^{-4}$\,mbar (film oxidation) during deposition.
The boron nitride heater was set to $790\,^\circ$C, which corresponds
approximately to $700\,^\circ$C at the substrate surface and a cooling rate of $10\,^\circ$C/min with 20\,min holding steps at 690, 490, and $290\,^\circ$C was used after deposition.

\subsection{Focussed ion beam preparation}

TEM samples were prepared in a Thermo Fisher Scientific Helios G4 gallium FIB. Electron and FIB deposited platinum was used to form a conductive protection bar on the surface of the sputtered platinum front contact. A cross-sectional slice was transferred to a DENSsolutions MEMS chip including two biasing and four heating contacts. After mounting on the chip, FIB platinum was used for electrical contacting of the STNO substrate as well as the front contact to the biasing contacts and vertical cuts were milled to avoid short circuits of the rectifying junction. 
The front and rear thinning patterns were rotated by 5\,$^\circ$ to create a wedge-shaped sample and the acceleration voltage during final ion polishing was set to 2\,kV.

\subsection{Transmission electron microscopy}

STEM-EBIC expriments were conducted in an FEI Titan 80–300 G2 ETEM operated at 300\,kV and using a DENSsolutions Lightning D7+ sample holder. The convergence semi-angle was set to 10\,mrad and the beam current to 42\,pA. STEM-EBIC currents were converted to a voltage signal with a Stanford Research Systems SR570 current preamplifier and recorded synchronously to annular dark-field signals as well as EELS data using a Gatan Quantum 965 ER image filter including an UltraScan 1000XP CCD camera and a collection semi-angle of 47\,mrad. The local sample thickness in the STNO substrate was determined via EELS using the log-ratio method and a mean free path of 149.1\,nm as suggested by the parametrization in \cite{iakoubovskii2008thickness}.

\medskip
\textbf{Supporting Information} \par 
Supporting Information is available from the Wiley Online Library or from the author.

\medskip
\textbf{Acknowledgements} \par 
The authors thank B. Kressdorf for providing the macroscopic heterojunction sample.
This work was financially supported by the Deutsche Forschungsgemeinschaft (DFG, German Research Foundation) 217133147/SFB 1073, projects B02, C02, and project 429413061 (SE560/-1) as well as by the German State of Lower Saxony (FuturePV project).
The use of equipment of the “Collaborative Laboratory and User Facility for Electron Microscopy” (CLUE, Göttingen) is gratefully acknowledged.

\medskip
\textbf{Author Contributions Statement}
T.M.:
conceptualization; 
data curation;
formal analysis;
investigation;
methodology;
software;
validation;
visualization;
writing—original draft;
writing—review \& editing.
C.F.:
formal analysis;
investigation;
methodology;
validation;
writing—review \& editing.
D.A.E.:
formal analysis;
investigation;
methodology;
software;
validation;
writing—review \& editing.
P.P.-P.:
conceptualization; 
formal analysis;
methodology;
software;
validation;
writing—review \& editing.
J.L.:
investigation;
methodology;
software;
writing—review \& editing.
C.J.:
writing—review \& editing;
resources;
supervision;
funding acquisition.
M.S.:
conceptualization; 
formal analysis;
methodology;
writing—review \& editing;
resources;
funding acquisition;
supervision;
project administration.

\medskip
\textbf{Data Availability Statement}

The data that support the findings of this study are openly available in \cite{publication_data} and additional raw data are available from the corresponding author upon reasonable request.

\medskip

%
\bibliographystyle{MSP}
\bibliography{bibliography}

@article{hackett1972electron,
  title={Electron-Beam Excited Minority-Carrier Diffusion Profiles in Semiconductors},
  author={Hackett Jr, WH},
  journal={Journal of Applied Physics},
  volume={43},
  number={4},
  pages={1649--1654},
  year={1972},
  publisher={American Institute of Physics}
}

@article{jastrzebski1975application,
  title={Application of scanning electron microscopy to determination of surface recombination velocity: GaAs},
  author={Jastrzebski, L and Lagowski, J and Gatos, HC},
  journal={Applied Physics Letters},
  volume={27},
  number={10},
  pages={537--539},
  year={1975},
  publisher={American Institute of Physics}
}

@article{van1950theory,
  title={Theory of the flow of electrons and holes in germanium and other semiconductors},
  author={Van Roosbroeck, W},
  journal={The Bell System Technical Journal},
  volume={29},
  number={4},
  pages={560--607},
  year={1950},
  publisher={Nokia Bell Labs}
}

@article{van1955injected,
  title={Injected current carrier transport in a semi-infinite semiconductor and the determination of lifetimes and surface recombination velocities},
  author={Van Roosbroeck, W},
  journal={Journal of Applied Physics},
  volume={26},
  number={4},
  pages={380--391},
  year={1955},
  publisher={American Institute of Physics}
}

@phdthesis{peretzki2019implementation,
  title={Implementation and quantification of scanning transmission EBIC experiments for measuring nanometer diffusion lengths in manganite-titanite pn heterojunctions},
  author={Peretzki, Patrick},
  year={2019},
  school={Nieders{\"a}chsische Staats-und Universit{\"a}tsbibliothek G{\"o}ttingen}
}

@article{haney2016depletion,
  title={Depletion region surface effects in electron beam induced current measurements},
  author={Haney, Paul M and Yoon, Heayoung P and Gaury, Benoit and Zhitenev, Nikolai B},
  journal={Journal of applied physics},
  volume={120},
  number={9},
  year={2016},
  publisher={AIP Publishing}
}

@article{iakoubovskii2008thickness,
  title={Thickness measurements with electron energy loss spectroscopy},
  author={Iakoubovskii, K and Mitsuishi, K and Nakayama, Y and Furuya, K},
  journal={Microscopy research and technique},
  volume={71},
  number={8},
  pages={626--631},
  year={2008},
  publisher={Wiley Online Library}
}

@article{massiot2020progress,
  title={Progress and prospects for ultrathin solar cells},
  author={Massiot, In{\`e}s and Cattoni, Andrea and Collin, St{\'e}phane},
  journal={Nature Energy},
  volume={5},
  number={12},
  pages={959--972},
  year={2020},
  publisher={Nature Publishing Group UK London}
}

@article{shockley32detailed,
  title={Detailed Balance Limit of Efficiency of pn Junction Solar Cells”(1961)},
  author={Shockley, W and Queisser, HJ},
  journal={J Appl Phys},
  volume={32},
  pages={510}
}

@article{queisser1998defects,
  title={Defects in semiconductors: some fatal, some vital},
  author={Queisser, Hans J and Haller, Eugene E},
  journal={Science},
  volume={281},
  number={5379},
  pages={945--950},
  year={1998},
  publisher={American Association for the Advancement of Science}
}

@article{ball2016defects,
  title={Defects in perovskite-halides and their effects in solar cells},
  author={Ball, James M and Petrozza, Annamaria},
  journal={Nature Energy},
  volume={1},
  number={11},
  pages={1--13},
  year={2016},
  publisher={Nature Publishing Group}
}

@article{green2014emergence,
  title={The emergence of perovskite solar cells},
  author={Green, Martin A and Ho-Baillie, Anita and Snaith, Henry J},
  journal={Nature photonics},
  volume={8},
  number={7},
  pages={506--514},
  year={2014},
  publisher={Nature Publishing Group UK London}
}

@article{wang2012electronics,
  title={Electronics and optoelectronics of two-dimensional transition metal dichalcogenides},
  author={Wang, Qing Hua and Kalantar-Zadeh, Kourosh and Kis, Andras and Coleman, Jonathan N and Strano, Michael S},
  journal={Nature nanotechnology},
  volume={7},
  number={11},
  pages={699--712},
  year={2012},
  publisher={Nature Publishing Group UK London}
}

@article{kressdorf2020room,
  title={Room-temperature hot-polaron photovoltaics in the charge-ordered state of a layered perovskite oxide heterojunction},
  author={Kressdorf, Birte and Meyer, Tobias and Belenchuk, Alexandr and Shapoval, O and Ten Brink, M and Melles, S and Ross, U and Hoffmann, J and Moshnyaga, Vasily and Seibt, Michael and others},
  journal={Physical Review Applied},
  volume={14},
  number={5},
  pages={054006},
  year={2020},
  publisher={APS}
}

@article{kressdorf2021orbital,
  title={Orbital-order phase transition in Pr 1- x Ca x MnO 3 probed by photovoltaics},
  author={Kressdorf, Birte and Meyer, T and Ten Brink, M and Seick, C and Melles, S and Ottinger, N and Titze, T and Meer, H and Weisser, A and Hoffmann, J and others},
  journal={Physical Review B},
  volume={103},
  number={23},
  pages={235122},
  year={2021},
  publisher={APS}
}

@inproceedings{meyer2019high,
  title={High-resolution scanning transmission EBIC analysis of misfit dislocations at perovskite Pn-heterojunctions},
  author={Meyer, T and Kressdorf, B and Lindner, J and Peretzki, P and Roddatis, Vladimir and Jooss, Christian and Seibt, Michael},
  booktitle={Journal of Physics: Conference Series},
  volume={1190},
  number={1},
  pages={012009},
  year={2019},
  organization={IOP Publishing}
}

@article{hoffmann2020fabrication,
  title={Fabrication of tin-based halide perovskites by pulsed laser deposition},
  author={Hoffmann-Urlaub, Sarah and Zhang, Yaodong and Wang, Zhaodong and Kressdorf, Birte and Meyer, Tobias},
  journal={Applied Physics A},
  volume={126},
  number={7},
  pages={553},
  year={2020},
  publisher={Springer}
}

@article{meyer2021phase,
  title={Phase Transitions in a Perovskite Thin Film Studied by Environmental In Situ Heating Nano-Beam Electron Diffraction},
  author={Meyer, Tobias and Kressdorf, Birte and Roddatis, Vladimir and Hoffmann, J{\"o}rg and Jooss, Christian and Seibt, Michael},
  journal={Small Methods},
  volume={5},
  number={9},
  pages={2100464},
  year={2021},
  publisher={Wiley Online Library}
}

@article{flathmann2024relationship,
  title={Relationship between structure and charge/orbital order in epitaxial single layer Ruddlesden--Popper manganite thin films},
  author={Flathmann, Christoph and Meyer, Tobias and Ross, Ulrich and Dehning, Annika and Jooss, Christian and Seibt, Michael},
  journal={APL Materials},
  volume={12},
  number={6},
  year={2024},
  publisher={AIP Publishing}
}

@article{wu2023toward,
  title={Toward accurate measurement of electromagnetic field by retrieving and refining the center position of non-uniform diffraction disks in Lorentz 4D-STEM},
  author={Wu, Lijun and Han, Myung-Geun and Zhu, Yimei},
  journal={Ultramicroscopy},
  volume={250},
  pages={113745},
  year={2023},
  publisher={Elsevier}
}

@article{flathmann2025sequential,
  title={Sequential tilting 4D-STEM for improved momentum-resolved STEM field mapping},
  author={Flathmann, Christoph and Ross, Ulrich and Belz, J{\"u}rgen and Beyer, Andreas and Volz, Kerstin and Seibt, Michael and Meyer, Tobias},
  journal={Microscopy and Microanalysis},
  volume={31},
  number={5},
  pages={ozaf086},
  year={2025},
  publisher={Oxford University Press US}
}

@article{flathmann2023composition,
  title={Composition and electronic structure of SiO x/TiO y/Al passivating carrier selective contacts on n-type silicon solar cells},
  author={Flathmann, Christoph and Meyer, Tobias and Titova, Valeriya and Schmidt, Jan and Seibt, Michael},
  journal={Scientific Reports},
  volume={13},
  number={1},
  pages={3124},
  year={2023},
  publisher={Nature Publishing Group UK London}
}

@article{fakharuddin2022perovskite,
  title={Perovskite light-emitting diodes},
  author={Fakharuddin, Azhar and Gangishetty, Mahesh K and Abdi-Jalebi, Mojtaba and Chin, Sang-Hyun and bin Mohd Yusoff, Abd Rashid and Congreve, Daniel N and Tress, Wolfgang and Deschler, Felix and Vasilopoulou, Maria and Bolink, Henk J},
  journal={Nature Electronics},
  volume={5},
  number={4},
  pages={203--216},
  year={2022},
  publisher={Nature Publishing Group UK London}
}

@article{sheen2022highly,
  title={Highly efficient blue InGaN nanoscale light-emitting diodes},
  author={Sheen, Mihyang and Ko, Yunhyuk and Kim, Dong-uk and Kim, Jongil and Byun, Jin-ho and Choi, YongSeok and Ha, Jonghoon and Yeon, Ki Young and Kim, Dohyung and Jung, Jungwoon and others},
  journal={Nature},
  volume={608},
  number={7921},
  pages={56--61},
  year={2022},
  publisher={Nature Publishing Group UK London}
}

@article{miyasaka2020perovskite,
  title={Perovskite solar cells: can we go organic-free, lead-free, and dopant-free?},
  author={Miyasaka, Tsutomu and Kulkarni, Ashish and Kim, Gyu Min and {\"O}z, Senol and Jena, Ajay K},
  journal={Advanced Energy Materials},
  volume={10},
  number={13},
  pages={1902500},
  year={2020},
  publisher={Wiley Online Library}
}

@article{du2024improving,
  title={Improving the Stability of Halide Perovskites for Photo-, Electro-, Photoelectro-Chemical Applications},
  author={Du, Fenqi and Liu, Xiaolong and Liao, Jinfeng and Yu, Dejian and Zhang, Nan and Chen, Yiwang and Liang, Chao and Yang, Shengchun and Fang, Guojia},
  journal={Advanced Functional Materials},
  volume={34},
  number={12},
  pages={2312175},
  year={2024},
  publisher={Wiley Online Library}
}

@article{lindner2024reconstruction,
  title={Reconstruction of Angstrom resolution exit-waves by the application of drift-corrected phase-shifting off-axis electron holography},
  author={Lindner, Jonas and Ross, Ulrich and Meyer, Tobias and Boureau, Victor and Seibt, Michael and Jooss, Ch},
  journal={Ultramicroscopy},
  volume={256},
  pages={113880},
  year={2024},
  publisher={Elsevier}
}

@article{schmidt2018surface,
  title={Surface passivation of crystalline silicon solar cells: Present and future},
  author={Schmidt, Jan and Peibst, Robby and Brendel, Rolf},
  journal={Solar Energy Materials and Solar Cells},
  volume={187},
  pages={39--54},
  year={2018},
  publisher={Elsevier}
}

@article{titova2018implementation,
  title={Implementation of full-area-deposited electron-selective TiOx layers into silicon solar cells},
  author={Titova, Valeriya and Schmidt, Jan},
  journal={AIP Advances},
  volume={8},
  number={12},
  year={2018},
  publisher={AIP Publishing}
}

@article{blakers2019development,
  title={Development of the PERC solar cell},
  author={Blakers, Andrew},
  journal={IEEE Journal of Photovoltaics},
  volume={9},
  number={3},
  pages={629--635},
  year={2019},
  publisher={IEEE}
}

@article{trupke2012photoluminescence,
  title={Photoluminescence imaging for photovoltaic applications},
  author={Trupke, T and Mitchell, B and Weber, JW and McMillan, W and Bardos, RA and Kroeze, R},
  journal={Energy Procedia},
  volume={15},
  pages={135--146},
  year={2012},
  publisher={Elsevier}
}

@article{coenen2017cathodoluminescence,
  title={Cathodoluminescence for the 21st century: Learning more from light},
  author={Coenen, T and Haegel, NM},
  journal={Applied Physics Reviews},
  volume={4},
  number={3},
  year={2017},
  publisher={AIP Publishing}
}

@article{vishwakarma2018direct,
  title={A direct measurement of higher photovoltage at grain boundaries in CdS/CZTSe solar cells using KPFM technique},
  author={Vishwakarma, Manoj and Varandani, Deepak and Andres, Christian and Romanyuk, Yaroslav E and Haass, Stefan G and Tiwari, Ayodhya N and Mehta, Bodh R},
  journal={Solar Energy Materials and Solar Cells},
  volume={183},
  pages={34--40},
  year={2018},
  publisher={Elsevier}
}

@article{matsumura2014characterization,
  title={Characterization of carrier concentration in CIGS solar cells by scanning capacitance microscopy},
  author={Matsumura, Koji and Fujita, Takaya and Itoh, Hiroshi and Fujita, Daisuke},
  journal={Measurement Science and Technology},
  volume={25},
  number={4},
  pages={044020},
  year={2014},
  publisher={IOP Publishing}
}

@article{eyben2011development,
  title={Development and optimization of scanning spreading resistance microscopy for measuring the two-dimensional carrier profile in solar cell structures},
  author={Eyben, Pierre and Seidel, Felix and Hantschel, Thomas and Schulze, Andreas and Lorenz, Anne and De Castro, Angel Uruena and Van Gestel, Dries and John, Joachim and Horzel, Joerg and Vandervorst, Wilfried},
  journal={physica status solidi (a)},
  volume={208},
  number={3},
  pages={596--599},
  year={2011},
  publisher={Wiley Online Library}
}

@article{moralejo2010lbic,
  title={LBIC and reflectance mapping of multicrystalline Si solar cells},
  author={Moralejo, B and Gonz{\'a}lez, MA and Jim{\'e}nez, J and Parra, V and Mart{\'\i}nez, O and Guti{\'e}rrez, J and Charro, O},
  journal={Journal of electronic materials},
  volume={39},
  number={6},
  pages={663--670},
  year={2010},
  publisher={Springer}
}

@article{leamy1982charge,
  title={Charge collection scanning electron microscopy},
  author={Leamy, HJ},
  journal={Journal of Applied Physics},
  volume={53},
  number={6},
  pages={R51--R80},
  year={1982},
  publisher={American Institute of Physics}
}

@article{donolato1983evaluation,
  title={Evaluation of diffusion lengths and surface recombination velocities from electron beam induced current scans},
  author={Donolato, C},
  journal={Applied Physics Letters},
  volume={43},
  number={1},
  pages={120--122},
  year={1983}
}

@article{donolato1999reconstruction,
  title={Reconstruction of the charge collection probability in a semiconductor device from the derivative of collection efficiency data},
  author={Donolato, C},
  journal={Applied Physics Letters},
  volume={75},
  number={25},
  pages={4004--4006},
  year={1999},
  publisher={American Institute of Physics}
}

@phdthesis{meyer2020structural,
  title={Structural and electronic investigation of strongly correlated transition metal oxide perovskite thin films and interfaces using in-situ transmission electron microscopy},
  author={Meyer, Tobias and others},
  year={2020},
  school={PhD thesis, Dissertation, G{\"o}ttingen, Georg-August Universit{\"a}t}
}

@article{conlan2021electron,
  title={Electron beam induced current microscopy of silicon p--n junctions in a scanning transmission electron microscope},
  author={Conlan, Aidan P and Moldovan, Grigore and Bruas, Lucas and Monroy, Eva and Cooper, David},
  journal={Journal of Applied Physics},
  volume={129},
  number={13},
  year={2021},
  publisher={AIP Publishing}
}

@article{shalimova2018random,
  title={Random walk on spheres method for solving anisotropic drift-diffusion problems},
  author={Shalimova, Irina and Sabelfeld, Karl K},
  journal={Monte Carlo Methods and Applications},
  volume={24},
  number={1},
  pages={43--54},
  year={2018},
  publisher={De Gruyter}
}

@article{dyck2023direct,
  title={Direct imaging of electron density with a scanning transmission electron microscope},
  author={Dyck, Ondrej and Almutlaq, Jawaher and Lingerfelt, David and Swett, Jacob L and Oxley, Mark P and Huang, Bevin and Lupini, Andrew R and Englund, Dirk and Jesse, Stephen},
  journal={Nature Communications},
  volume={14},
  number={1},
  pages={7550},
  year={2023},
  publisher={Nature Publishing Group UK London}
}

@article{mecklenburg2019electron,
  title={Electron beam-induced current imaging with two-angstrom resolution},
  author={Mecklenburg, Matthew and Hubbard, William A and Lodico, Jared J and Regan, BC},
  journal={Ultramicroscopy},
  volume={207},
  pages={112852},
  year={2019},
  publisher={Elsevier}
}

@article{duchamp2020stem,
  title={STEM electron beam-induced current measurements of organic-inorganic perovskite solar cells},
  author={Duchamp, Martial and Hu, H and Lam, Yeng Ming and Dunin-Borkowski, RE and Boothroyd, Christopher B},
  journal={Ultramicroscopy},
  volume={217},
  pages={113047},
  year={2020},
  publisher={Elsevier}
}

@article{cabanel2006low,
  title={Low temperature semi-quantitative analysis of local electrical field in silicon diode by transmission electron microscopy},
  author={Cabanel, C and Brouri, D and Laval, JY},
  journal={The European Physical Journal-Applied Physics},
  volume={34},
  number={2},
  pages={107--116},
  year={2006},
  publisher={EDP Sciences}
}

@article{luke1985quantification,
  title={Quantification of the effects of generation volume, surface recombination velocity, and diffusion length on the electron-beam-induced current and its derivative: Determination of diffusion lengths in the low micron and submicron ranges},
  author={Luke, Keung L and von Roos, Oldwig and Cheng, Li-jen},
  journal={Journal of applied physics},
  volume={57},
  number={6},
  pages={1978--1984},
  year={1985},
  publisher={American Institute of Physics}
}

@article{ong1994direct,
  title={A direct and accurate method for the extraction of diffusion length and surface recombination velocity from an EBIC line scan},
  author={Ong, VKS and Phang, JCH and Chan, DSH},
  journal={Solid-state electronics},
  volume={37},
  number={1},
  pages={1--7},
  year={1994},
  publisher={Elsevier}
}

@article{berz1976theory,
  title={Theory of life time measurements with the scanning electron microscope: Steady state},
  author={Berz, F and Kuiken, HK},
  journal={Solid-State Electronics},
  volume={19},
  number={6},
  pages={437--445},
  year={1976},
  publisher={Elsevier}
}

@article{schneider2025stem,
  title={STEM EBIC as a Quantitative Probe of Semiconductor Devices},
  author={Schneider, Sebastian and Beckert, Sebastian and Hammer, Ren{\'e} and K{\"o}nig, Markus and Moldovan, Grigore and Pohl, Darius},
  journal={arXiv preprint arXiv:2511.11528},
  year={2025}
}

@misc{hubbard2023emission,
  title={Emission-Based Temperature Mapping with STEM EBIC},
  author={Hubbard, William A and Mecklenburg, Matthew and Chan, Ho Leung and Regan, BC},
  year={2023},
  publisher={Oxford University Press US}
}

@article{zutter2021mapping,
  title={Mapping Charge Recombination and the Effect of Point-Defect Insertion in GaAs Nanowire Heterojunctions},
  author={Zutter, Brian T and Kim, Hyunseok and Hubbard, William A and Ren, Dingkun and Mecklenburg, Matthew and Huffaker, Diana and Regan, BC},
  journal={Physical Review Applied},
  volume={16},
  number={4},
  pages={044030},
  year={2021},
  publisher={APS}
}

@article{tan2013study,
  title={The study of the charge collection of the normal-collector configuration},
  author={Tan, Chee Chin and Ong, Vincent KS and Radhakrishnan, K},
  journal={Progress in Photovoltaics: Research and Applications},
  volume={21},
  number={5},
  pages={986--995},
  year={2013},
  publisher={Wiley Online Library}
}

@article{vlasov2023secondary,
  title={Secondary electron induced current in scanning transmission electron microscopy: an alternative way to visualize the morphology of nanoparticles},
  author={Vlasov, Evgenii and Skorikov, Alexander and S{\'a}nchez-Iglesias, Ana and Liz-Marzan, Luis M and Verbeeck, Johan and Bals, Sara},
  journal={ACS Materials Letters},
  volume={5},
  number={7},
  pages={1916--1921},
  year={2023},
  publisher={ACS Publications}
}

@article{rau1999two,
  title={Two-dimensional mapping of the electrostatic potential in transistors by electron holography},
  author={Rau, WD and Schwander, P and Baumann, FH and H{\"o}ppner, W and Ourmazd, A},
  journal={Physical Review Letters},
  volume={82},
  number={12},
  pages={2614},
  year={1999},
  publisher={APS}
}

@article{twitchett2002quantitative,
  title={Quantitative electron holography of biased semiconductor devices},
  author={Twitchett, AC and Dunin-Borkowski, RE and Midgley, PA},
  journal={Physical review letters},
  volume={88},
  number={23},
  pages={238302},
  year={2002},
  publisher={APS}
}

@article{peretzki2017low,
  title={Low energy scanning transmission electron beam induced current for nanoscale characterization of p--n junctions},
  author={Peretzki, Patrick and Ifland, Benedikt and Jooss, Christian and Seibt, Michael},
  journal={physica status solidi (RRL)--Rapid Research Letters},
  volume={11},
  number={1},
  pages={1600358},
  year={2017},
  publisher={Wiley Online Library}
}

@article{toyama2023real,
  title={Real-space observation of a two-dimensional electron gas at semiconductor heterointerfaces},
  author={Toyama, Satoko and Seki, Takehito and Kanitani, Yuya and Kudo, Yoshihiro and Tomiya, Shigetaka and Ikuhara, Yuichi and Shibata, Naoya},
  journal={Nature Nanotechnology},
  volume={18},
  number={5},
  pages={521--528},
  year={2023},
  publisher={Nature Publishing Group UK London}
}

@article{wartelle2025sub,
  title={Sub-microradian angular detection limits for field mapping by Lorentz 4D scanning transmission electron microscopy on a Si p--n junction},
  author={Wartelle, Alexis and da Silva, Bruno C and Cooper, David and den Hertog, Martien I},
  journal={Journal of Applied Physics},
  volume={138},
  number={10},
  year={2025},
  publisher={AIP Publishing}
}

@article{anada2020direct,
  title={Direct visualization of the photovoltaic effect in a single-junction GaAs cell via in situ electron holography},
  author={Anada, Satoshi and Hirayama, Tsukasa and Sasaki, Hirokazu and Yamamoto, Kazuo},
  journal={Journal of Applied Physics},
  volume={128},
  number={24},
  year={2020},
  publisher={AIP Publishing}
}

@article{ccelik2024simple,
  title={A simple and intuitive model for long-range 3D potential distributions of in-operando TEM-samples: Comparison with electron holographic tomography},
  author={{\c{C}}elik, H{\"u}seyin and Fuchs, Robert and Gaebel, Simon and G{\"u}nther, Christian M and Lehmann, Michael and Wagner, Tolga},
  journal={Ultramicroscopy},
  volume={267},
  pages={114057},
  year={2024},
  publisher={Elsevier}
}

@article{denaix2023inversion,
  title={Inversion of the internal electric field due to inhomogeneous incorporation of Ge dopants in GaN/AlN heterostructures studied by off-axis electron holography},
  author={Denaix, Lou and Castioni, Florian and Bryan, Matthew and Cooper, David and Monroy, Eva},
  journal={ACS Applied Materials \& Interfaces},
  volume={15},
  number={8},
  pages={11208--11215},
  year={2023},
  publisher={ACS Publications}
}

@data{publication_data,
author = {Meyer, Tobias and Flathmann, Christoph and Ehrlich, David A. and Paap-Peretzki, Patrick and Lindner, Jonas and Jooß, Christian and Seibt, Michael},
publisher = {GRO.data},
title = {{Data for: Quantitative modeling of excess charge carrier diffusion and recombination in electron transparent samples}},
year = {2026},
version = {V1},
doi = {10.25625/IWVEOW},
url = {https://doi.org/10.25625/IWVEOW}
}



\newpage

\begin{figure}
\textbf{Table of Contents}\\
\medskip
  \includegraphics[width=\linewidth]{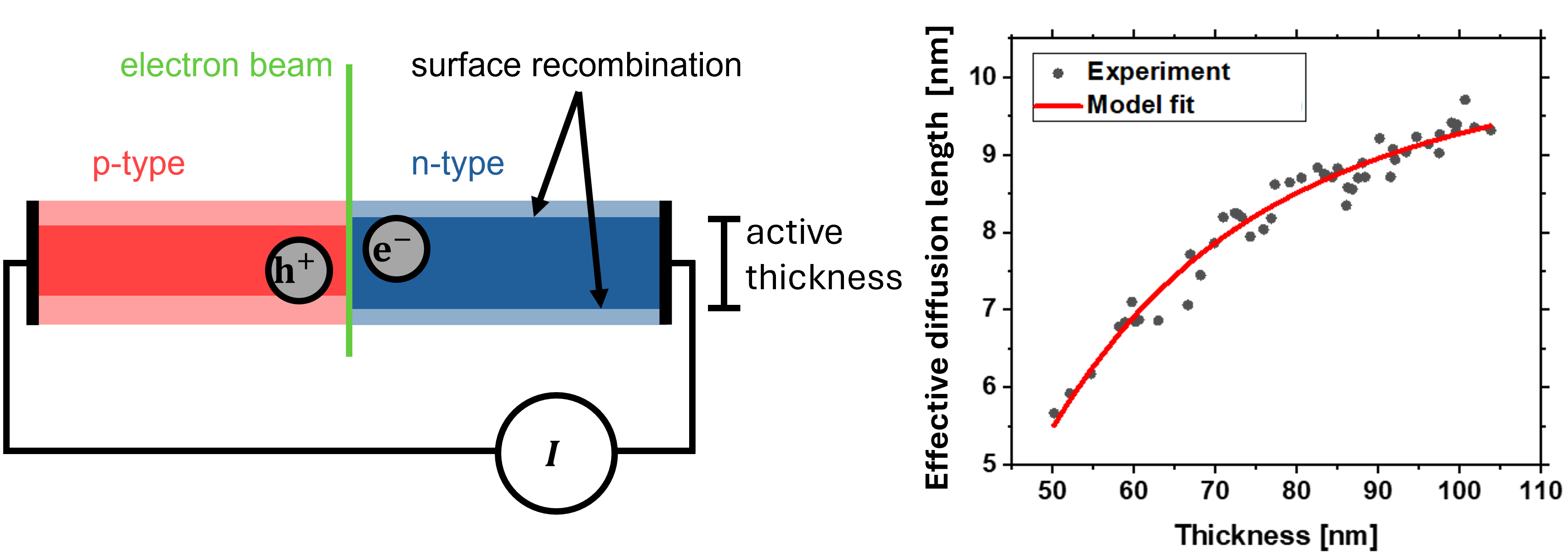}
  \raggedright{Diffusion and recombination of excess charge carriers in a semiconducting material are considered using a finite geometry that is tailored for electron beam induced current measurements in a scanning transmission electron microscope. A quantitative model is developed based on analytical considerations as well as finite element simulations and applied successfully to experimental data.}
\end{figure}

\newpage
~
\newpage
\setcounter{page}{1}
\subsection*{Supporting Information to}
\sectionmark{}

\section*{Quantitative models for excess carrier diffusion and recombination in STEM-EBIC experiments on semiconductor nanostructures}
\author{Tobias Meyer*}
\author{Christoph Flathmann}
\author{David Alexander Ehrlich}
\author{Patrick Paap-Peretzki}
\author{Jonas Lindner}
\author{Christian Jooß}
\author{Michael Seibt} 

\subsection*{Details of FEM simulations}

The parameters used in the FEM simulations are summarized in Table S1.
The boundary conditions corresponding to the boundaries indexed in Figure S1
and using the parameters listed in Table S1 are given in Table S2
and STEM-EBIC currents for the geometry in Figure S1b were evaluated by integrating the total current density $j_\text{tot}$ at boundary 2 defined by:
\begin{align*}
j_\text{tot}=e\left(D_\text{e}\vec\nabla n-D_\text{h}\vec\nabla p\right)\,.
\end{align*}
No significant differences between currents extracted at boundary 2 and 3 were observed.

\begin{table}[!h]
 \raggedright{Table S1: Parameters used in the FEM simulations. If applicable, different values for the p- and n-type material used in the geometry shown in Figure \ref{fig:geometry}a are given in separate columns.}
  \begin{tabular}{@{}llcc@{}}
    \hline
    Quantity & Description & n-type material & p-type material\\
    \hline
$d$& Material width &  \multicolumn{2}{c}{1200\,nm}  \\
$T$ & Temperature &  \multicolumn{2}{c}{300\,K}  \\
$E_\text{g}$& Band gap &  \multicolumn{2}{c}{1.12\,eV}  \\
$\epsilon_\text{r}$&  Relative permittivity &  \multicolumn{2}{c}{11.9}  \\
$\mu_\text{e}$& Electron mobility &  \multicolumn{2}{c}{1100\,cm$^2$/V/s}  \\
$\mu_\text{h}$& Hole mobility &  \multicolumn{2}{c}{200\,cm$^2$/V/s}  \\
$D_\text{e}$& Electron diffusivity &  \multicolumn{2}{c}{0.0028437 m²/s}  \\
$D_\text{h}$& Hole diffusivity &  \multicolumn{2}{c}{5.1704E-4 m²/s}  \\
$N_\text{d}$& Donor concentration & 0\,cm$^{-3}$ & $10^{17}\,$cm$^{-3}$ \\
$N_\text{a}$& Acceptor concentration & $10^{17}\,$cm$^{-3}$ & 0\,cm$^{-3}$ \\
$N_\text{C}$& Conduction band effective DOS &  \multicolumn{2}{c}{$2.5055\times10^{17}\,$cm$^{-3}$}  \\
$N_\text{V}$& Valence band effective DOS &  \multicolumn{2}{c}{$2.5055\times10^{17}\,$cm$^{-3}$}  \\
$n_\text{i}$& Intrinsic charge carrier concentration &  \multicolumn{2}{c}{$9.802\times10^{9}\,$cm$^{-3}$}  \\
$n_1$& Mid-gap electron concentration &  \multicolumn{2}{c}{$9.802\times10^{9}\,$cm$^{-3}$}  \\
$p_1$& Mid-gap hole concentration &  \multicolumn{2}{c}{$9.802\times10^{9}\,$cm$^{-3}$}  \\
$n_\text{eq}$ & Equilibrium electron concentration & $10^{17}\,$cm$^{-3}$ & $9.6079\times10^{2}\,$cm$^{-3}$\\
$p_\text{eq}$ & Equilibrium hole concentration & $9.6079\times10^{2}\,$cm$^{-3}$ & $10^{17}\,$cm$^{-3}$\\
$V_\text{bi}$& Built-in potential &  \multicolumn{2}{c}{0.8344\,V}  \\
$r$& Recombination function &  \multicolumn{2}{c}{$B\cdot(np-n_\text{i}^2)$}  \\
$B$& Bulk recombination coefficient &  \multicolumn{2}{c}{$2.5\times10^{-6}\,$cm$^{3}$/s}  \\
$g(x,z)$& Generation function &  \multicolumn{2}{c}{$G_\text{power}/\sqrt{2\pi}/\sigma_\text{b}/E_\text{g}\cdot\exp{(-(x-x_\text{b})^2/2/\sigma_\text{b})}$}  \\
$\sigma_\text{b}$& Electron beam size &  \multicolumn{2}{c}{1\,nm}  \\
$G_\text{power}$& Electron beam strength &  \multicolumn{2}{c}{20\,W/cm}  \\
     \hline
  \end{tabular}
  \label{tab:simulation_parameters}
\end{table}

\begin{figure}[!h]
  \includegraphics[width=\linewidth]{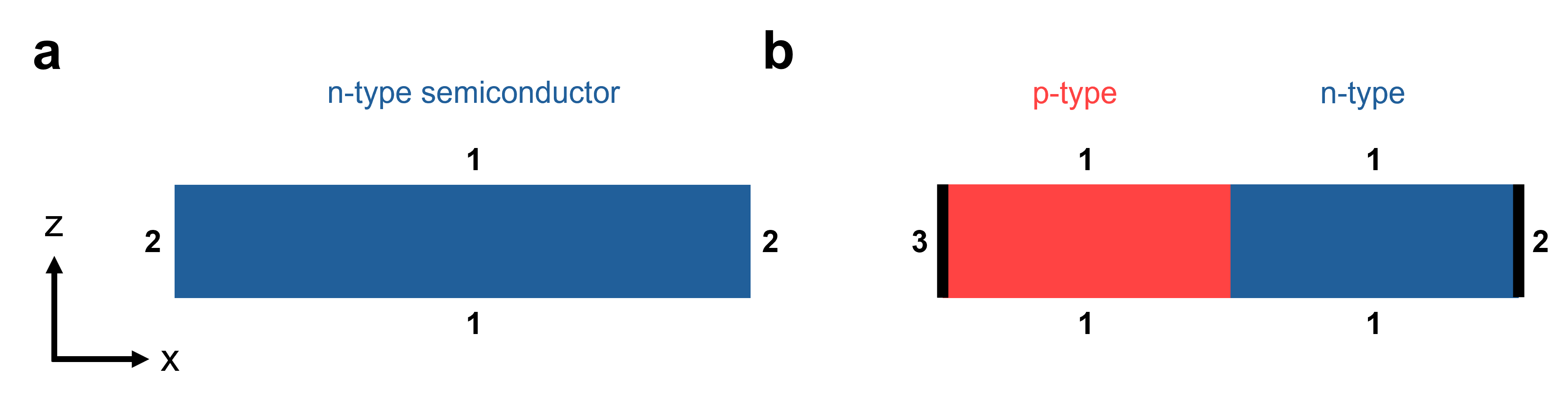}
  \raggedright{Figure S1: Boundary indices in the simulation models of the neutral n-type semiconductor (a) and the pn junction (b).}
  \label{fig:SI1}
\end{figure}

\begin{table}[!h]
 \raggedright{Table S2: Boundary conditions used in the FEM simulations.}
  \begin{tabular}{@{}llll@{}}
    \hline
    Boundary index & Variable & Boundary condition type & Expression\\
    \hline
    1&Electric potential $\phi$&Neumann&$\vec{n}\vec\nabla\phi=0$\\
    1&Electron concentration $n$&Dirichlet&$n=n_\text{eq}$\\
    1&Hole concentration $p$&Dirichlet&$p=p_\text{eq}$\\
    2&Electric potential $\phi$&Dirichlet&$\phi=+V_\text{bi}/2$\\
    2&Electron concentration $n$&Neumann&$\hat{n}\cdot D\vec\nabla n=s\frac{np-n_i^2}{n+n_1+p+p_1}$\\
    2&Hole concentration $p$&Neumann&$\hat{n}\cdot D\vec\nabla p=s\frac{np-n_i^2}{n+n_1+p+p_1}$\\
    3&Electric potential $\phi$&Dirichlet&$\phi=-V_\text{bi}/2$\\
    3&Electron concentration $n$&Neumann&$\hat{n}\cdot D\vec\nabla n=s\frac{np-n_i^2}{n+n_1+p+p_1}$\\
    3&Hole concentration $p$&Neumann&$\hat{n}\cdot D\vec\nabla p=s\frac{np-n_i^2}{n+n_1+p+p_1}$\\
     \hline
  \end{tabular}
  \label{tab:boundary_conditions}
\end{table}

\newpage

\subsection*{Extraction of decay lengths via exponential fits}

To extract the decay length $L_\text{decay}$ in excess hole concentration and STEM-EBIC profiles, least-square fits to the function 
\begin{align*}
a\cdot\exp{(-x/L_\text{decay})}+c    
\end{align*}
were performed. Special care needs to be taken that data points too close to the contact/model boundary or space charge region -- with expected deviations from an exponential behavior -- are excluded from the fit.
In the FEM simulation, the distance of the right-most point in the fit interval to the contact/model boundary was set to 300\,nm, which is well above five times the bulk diffusion length of the modeled n-type material. In the experiment, the back contact is several micrometers apart from the scanning region and no data points were excluded on the right side of the STEM-EBIC profiles. On the other hand, the left-most point was varied and for each position an exponential fit was performed to extract both a value for the decay length and its standard deviation (obtained from the covariance matrix of the least-squares parameter estimates). The typical dependence of both is plotted in combination with the corresponding STEM-EBIC profiles for simulated and experimental data in Figure S2.
The value with the lowest standard deviation is chosen as the decay length as the related left-most point includes as many data points as possible, but excludes points too close to the space charge region causing an overestimation of the decay length.

\begin{figure}[!h]
  \includegraphics[width=\linewidth]{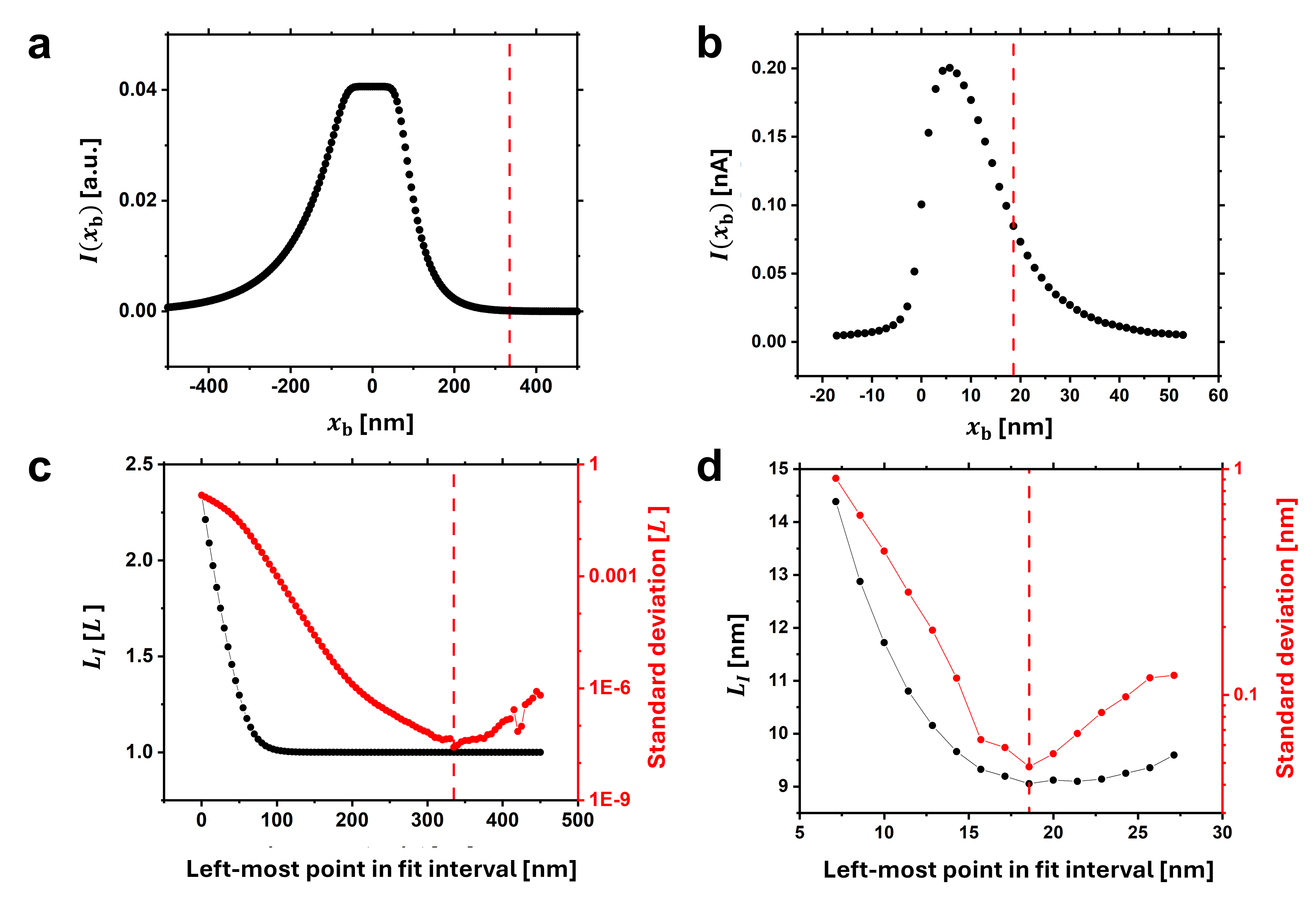}
  \raggedright{Figure S2: Examples of STEM-EBIC profiles including (a) simulated and (b) experimental data. The corresponding extracted decay length and its standard deviation obtained by an exponential fit is shown with respect to the left-most point in the fit interval in (c) and (d). The dashed red lines mark the points with minimal standard deviation and the related decay lengths were used as the values presented in the main text.}
  \label{fig:SI2}
\end{figure}

\newpage~\newpage

\subsection*{Local thickness determination}

The local sample thickness was determined with EELS using the log-ratio method and a mean free path of 149.1\,nm. Since the obtained values differ only by a few nanometer along the vertical direction, the thickness is evaluated along the red arrow in Figure S3a and assumed to be constant for vertical line profiles. The resulting values as a function of the horizontal position are shown in Figure S3b. The average increase in thickness corresponds to a wedge angle of 4.4\,$^\circ$, which is close to the 5.0\,$^\circ$ rotation angle set between front and rear thinning patterns in the FIB.

\begin{figure}[!h]
  \includegraphics[width=\linewidth]{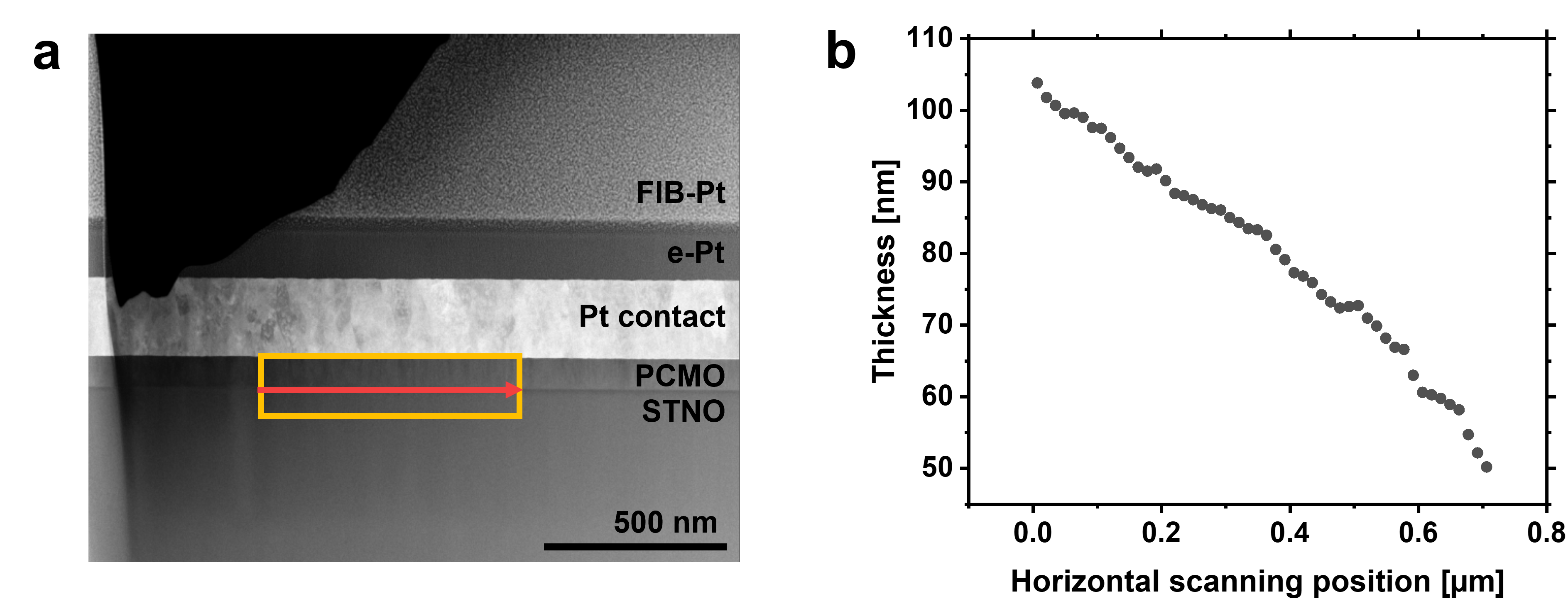}
  \raggedright{Figure S3: (a) Annular dark-field image including the experimentally investigated region of interest (yellow rectangle). The local thickness as a function of the horizontal position was extracted along the red arrow and is presented in (b).}
  \label{fig:SI3}
\end{figure}

\end{document}